\documentclass{aa}
\usepackage{natbib}
\usepackage{psfig,graphicx,epsfig}

\def\gsim{\;\lower4pt\hbox{${\buildrel\displaystyle >\over\sim}$}\,}
\def\lsim{\;\lower4pt\hbox{${\buildrel\displaystyle <\over\sim}$}\,}

\def \xmm {{\em XMM-Newton}}

\def \hcm {\hbox {\ifmmode $ atom cm$^{-2}\else atom cm$^{-2}$\fi}}
\def \arcmin {\hbox{$^\prime$}}
\def \arcsec {\hbox{$^{\prime\prime}$}}

\def\approxgt{\mathrel{\hbox{\rlap{\lower.55ex \hbox {$\sim$}}
        \kern-.3em \raise.4ex \hbox{$>$}}}}
\def\approxlt{\mathrel{\hbox{\rlap{\lower.55ex \hbox {$\sim$}}
        \kern-.3em \raise.4ex \hbox{$<$}}}}

\newcommand\U[1]{{\,\rm #1}}

\newcommand\E[1]{\times10^{#1}}

\newcommand\rs[1]{_\mathrm{#1}}

\newcommand\tht{\theta}
\newcommand\sg{\sigma}
\newcommand\Om{\Omega}

\newcommand\al{\alpha}

\def\lsim{\;\raise0.3ex\hbox{$<$\kern-0.75em\raise-1.1ex\hbox{$\sim$}}\;}
\def\gsim{\;\raise0.3ex\hbox{$>$\kern-0.75em\raise-1.1ex\hbox{$\sim$}}\;}

\def\beq{\begin{equation}}
\def\enq{\end{equation}}
\def\begar{\begin{eqnarray}}
\def\endar{\end{eqnarray}}
\def\mathnew{\mathsurround=0pt}
\def\simov#1#2{\lower .5pt\vbox{\baselineskip0pt \lineskip-.5pt
        \ialign{$\mathnew#1\hfil##\hfil$\crcr#2\crcr\sim\crcr}}}

\def \suz {{\it SUZAKU}}
\def \xmm {{\it XMM-Newton}}
\def \src {G54.1+0.3}
\def \epic {EPIC}

\newcommand\ord{{\cal O}}

\newcommand\tausca{\tau\rs{sca}}
\newcommand\Isca{I\rs{sca}}
\newcommand\Iintr{I\rs{intr}}
\newcommand\xmin{x\rs{min}}
\newcommand\xmax{x\rs{max}}

\newcommand\amax{a\rs{max}}
\newcommand\thtscal{\tht\rs{scal}}
\newcommand\thtn{\tht\rs{n}}

\newcommand\dsgdOmd{\frac{d\sg}{d\Om}}

\newcommand\zmin{z\rs{min}}
\newcommand\zmax{z\rs{max}}
\newcommand\CHANGED[1]{{\bf #1}}

\begin{document}


\title{\xmm\  and \suz\  detection of an X-ray emitting shell around the pulsar wind nebula \src}


\author{F. Bocchino\inst{1}
\and R. Bandiera\inst{2}
\and J. Gelfand\thanks{NSF Astronomy and Astrophysics Postdoctoral Fellow}\inst{3}
}

\offprints{e-mail: bocchino$@$astropa.inaf.it}

\institute{
       INAF-Osservatorio Astronomico di Palermo, Piazza del Parlamento 1,
       90134 Palermo, Italy
\and
       INAF-Osservatorio Astrofisico di Arcetri, Largo Enrico Fermi 1
       Firenze, Italy
\and
       Center for Cosmology and Particle Physics, New York University,
       4 Washington Place, New York, NY 10003, USA
}

\date{Received  / Accepted }

\authorrunning{Bocchino et al.}
\titlerunning{The shell of \src}

\abstract
  {Recent X-ray observations have proved to be very effective in detecting
  previously unknown supernova remnant shells around pulsar wind nebulae (PWNe),
  and in these cases the characteristics of the shell provide further
  clues on the evolutionary stage of the embedded PWN. However, it is
  not clear why some PWNe are still ``naked''. }
  {We carried out an X-ray observational campaign targeted at the PWN
  \src, the ``close cousin'' of the Crab, with the aim to detect the
  associated SNR shell.}
  {We analyzed an XMM-Newton and Suzaku observations of \src\  and we
  model out the contribution of dust scattering halo.}
  {We detected an intrinsic faint diffuse X-ray emission surrounding
  the PWN up to $\sim 6\arcmin$ ($\sim 10$ pc) from the pulsar, characterized by a
  hard spectrum, which can be modeled either with a power-law ($\gamma=
  2.9$) or with a thermal plasma model ($kT=2.0$ keV.)}
  {If the shell is thermal, we derive an explosion energy $E=0.5-1.6\times
  10^{51}$ erg, a pre-shock ISM density of 0.2 cm$^{-3}$ and an age
  of $\sim 2000$ yr. Using these results in the MHD model of PWN-SNR
  evolution, we obtain an excellent agreement between the predicted and
  observed location of the shell and PWN shock.}

\keywords{ISM: supernova remnants --
         (ISM:) dust, extinction, X-rays --
	 ISM, X-rays: individuals: \src}

\maketitle

\section{Introduction}

One of the most intriguing problems in the field of the Pulsar Wind
Nebulae (PWNe) study is the lack of a shell around some of these
objects. This is somehow disturbing for the consolidated picture of a
remnant of a core-collapse supernova, which indicates that the PWN is
expanding inside the host supernova remnant, giving rise to
a variety of complex phenomena, like reverberation, Rayleigh-Taylor
instability at the interface between the PWN and ejecta, rejuvenating
the shell, etc. (e.g. \citealt{vag01}; \citealt{bcf01}; \citealt{rac05};
\citealt{gsz09} and references therein). One of the reasons could be
the lack of deep observations aimed at the PWN surroundings. Indeed,
recently a shell-like component has been observed in many objects, such as
G21.5--0.9 (\citealt{bb04}; \citealt{bvc05}), G0.9+0.1 (\citealt{pdw03}),
3C58 (\citealt{bwm01}; \citealt{ghn07}).  Therefore,
X-ray observations are very effective for the discovery of
associated shell components, even in the presence of high absorption
column densities (G21.5--0.9 has $N_H\sim2\E{22}\U{cm^{-2}}$; G0.9+0.1
even $\sim10^{23}\U{cm^{-2}}$).

The objects in which pulsar, plerion, and shell are all detected
(collectively known as composite SNRs) are extremely important to set
the physical conditions for their modeling.  The properties and the
evolution of a PWN are determined by the interaction
of the pulsar wind with the ambient medium.  The effectiveness by
which the surrounding matter confines the PWN is very important to
determine the the level of the synchrotron emission from the nebula.
Therefore, measuring density and pressure in the shell component is
needed for better constraining the models of the PWN.
Moreover, the pulsar, the PWN and the shell would allow us to estimate in
independent ways some quantities, such as the actual age of the object,
or the internal pressure of the nebula.  This redundancy would allow us
also to verify our assumptions, like that on the level of equipartition
in the PWN and that on how reliable is the age estimated
from the pulsar spin-down properties.

\src\  is the Galactic PWN that most closely resembles
the Crab Nebula: this is the reason why \citet{lwa02} have dubbed it
``a close cousin of the Crab Nebula''.  Using Chandra data, \citet{lwa02}
have shown the presence of a well defined torus of $\sim10''$ in diameter,
together with elongations, toward E and W directions, which could be
ascribed to X-ray jets.  From those data, the size of the X-ray nebula
appears $\sim1'$, but the outer part of the nebula is very faint, and its
edge is poorly defined.  At radio wavelengths, instead, the nebular size
is $\sim1.5'$ (\citealt{vb88}), corresponding to $\sim2.7\,d_{6.2}$~pc
where $d_{6.2}$ is the distance of G54.1+0.3 in units of that estimated by
\citet{ltw08}, namely $d_{6.2}=d/6.2^{+1.0}_{-0.6}$ kpc.  It is important
to understand to which extent this difference in size is real (i.e.\ due to
synchrotron losses of the emitting electrons), or it is an artifact of
the limited X-ray sensitivity.

Radio maps show a rather amorphous structure, but the radio emission from
G54.1+0.3 is highly polarized, up to 20--30\% (\citealt{vb88}), and this
indicates (similarly to the case of the Crab Nebula) that the nebular
field is highly ordered.  The X-ray spectrum is a power law with a photon
index $-1.9$, an absorption column density $N_H\sim  1.6\E{22}\U{cm^{-2}}$
and an X-ray luminosity $L_X\sim  1.3\E{33}d_{6.2}^2\U{erg\,s^{-1}}$.
\citet{clb02} have detected the pulsar PSR J1930+1852 at the center of
the nebula, which has a period of 136 ms, a characteristic age of 2900
yr and a spin-down luminosity of $1.2\times 10^{37}$ erg s$^{-1}$.

Recently, \src\  has attracted some interest because \citet{kml08} have
identified an IR shell surrounding the PWN at a distance of $\sim 1.5'$
from the pulsar. The shell contains a dozen of IR compact sources.
\citet{kml08} suggests that the sources are young stellar objects,
whose formation has  been triggered by the wind of the progenitor of the SN.
This intriguing possibility has been questioned by \citet{tsr10},
who pointed out that the IR shell may be ejecta dust, rather than a
pre-existing ISM dense cloud.  \citet{ltw08} reported the presence
of a molecular cloud partially interacting with the PWN, on the basis
of the CO emission around the nebula.  Therefore, even if there is no
hint of a radio shell around this PWN, there are a number of evidences
for interaction between the PWN and the surroundings, so it is worth
searching for an X-ray shell.

In Sect. 2 we present a deep
X-ray campaign aimed to this PWN, which led to the detection of such a
shell. In Sect. 3 we estimate the contribution of the dust scattering
halo, showing that it is negligible at the shell location, and in Sect. 4
we discuss our findings comparing them to a PWN-SNR evolution model.

\section{Deep \xmm\  and \suz\  observations}

\begin{table}
  \caption{X-ray observations used in this work}
\label{obs}
\medskip
\centering\begin{minipage}{8.7cm}
\begin{tabular}{lcccc} \hline
  Satellite & ID & $T_{exp}$ (ks) & Obs. date \\ \hline

  \xmm  & 0406730101 & 41\footnote{PN exposure time after proton flare screening} &  26.09.2006\\
\suz  & 502077010 & 84 & 30.10.2007 \\
\hline
\end{tabular}
\end{minipage}
\end{table}

\begin{figure}
  \centerline{\psfig{file=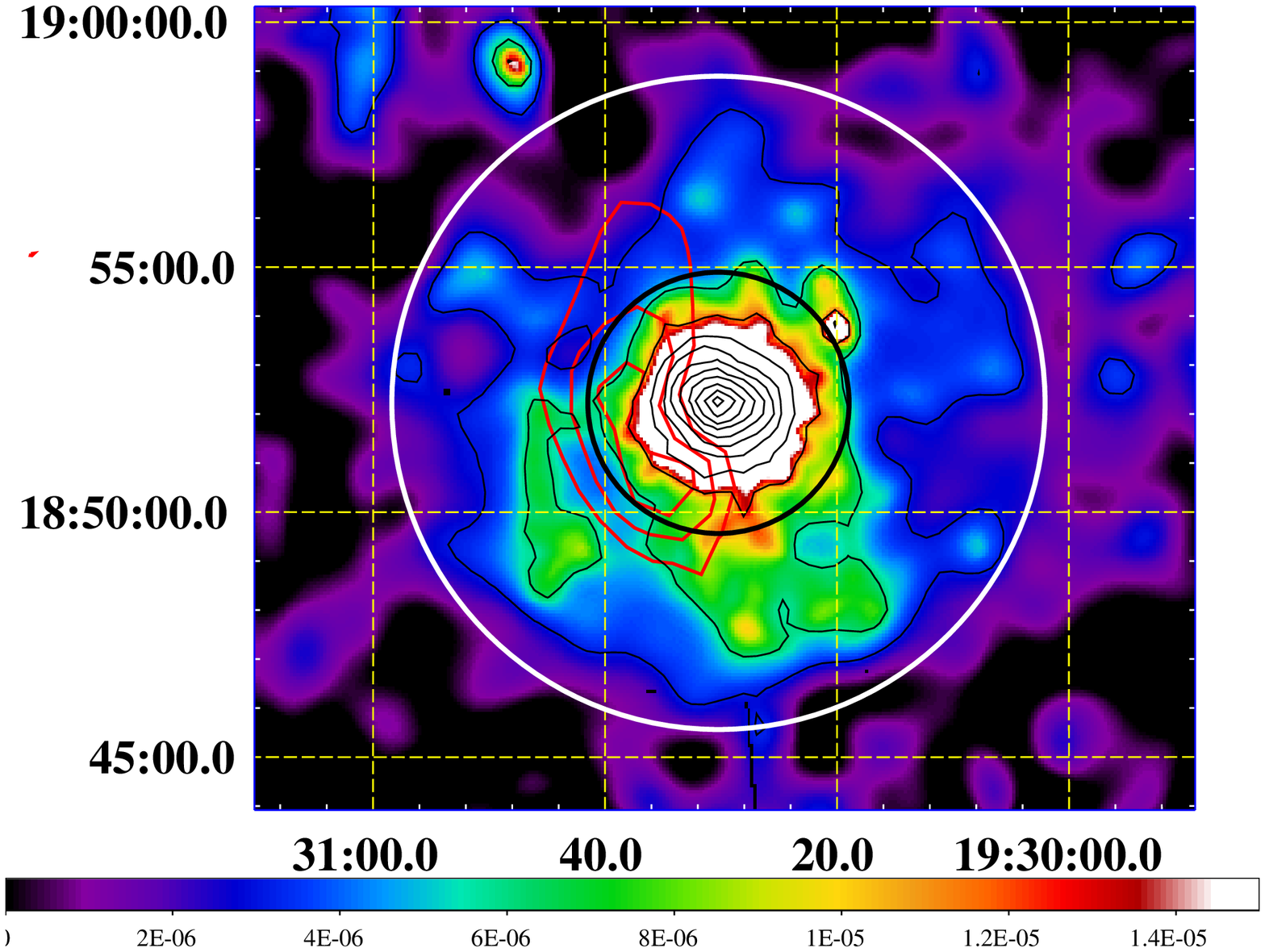,width=9.0cm}}
  \centerline{\psfig{file=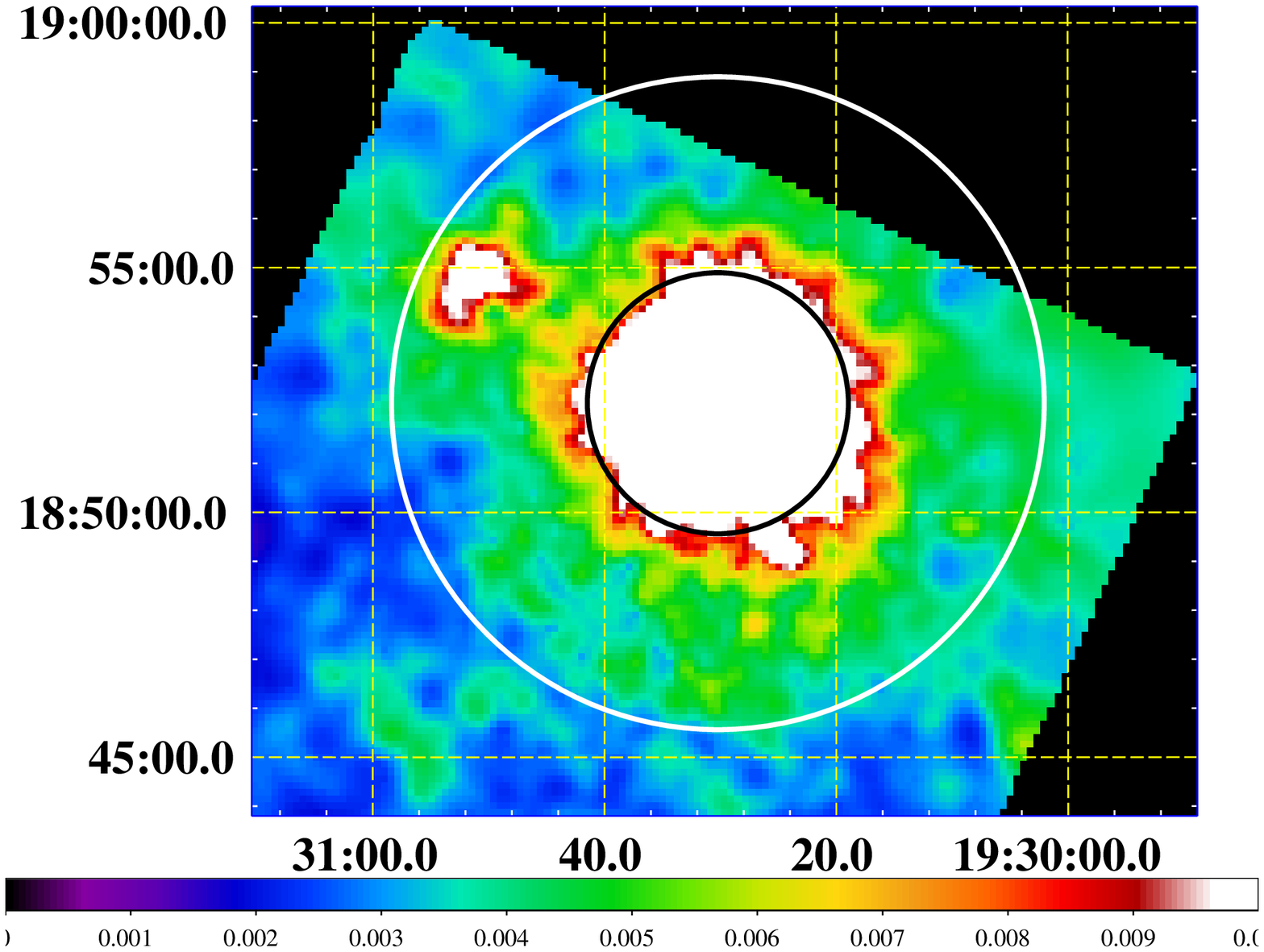,width=9.0cm}}
  \caption{{\em Top panel:} \xmm\  \epic\  image in the 1--7 keV energy
  band. The image is a mosaic of background subtracted and vignetting
  corrected images obtained with the PN, MOS1 and MOS2 cameras. The color
  scale has been chosen to maximize the visibility of the faint diffuse
  emission around the bright PWN. The inner (black) circle has a radius of
  $160^{\prime\prime}$ and includes the extension of the radio plerion,
  while the outer (white) circle has a radius of 400'' and is drawn to guide
  the eye on the extension of the shell. Black thin contours marks an
  increase of factor 2 in surface brightness from $3\times 10^{-6}$
  to $3.2\times 10^{-3}$ cnt s$^{-1}$ pix$^{-1}$, where 1 pixel is
  $4^{\prime\prime}\times 4^{\prime\prime}$. Red contours are drawn
  form the map of the 53 km s$^{-1}$ CO emission published in Fig. 3b
  of \protect\citet{ltw08}, and represent a molecular cloud possibly
  interacting with the PWN. {\em Bottom panel:} \suz\  image in the 1--7
  keV energy band. The image is a mosaic of background subtracted and
  vignetting corrected images obtained with the XIS0, XIS2 and XIS3. The
  field of view and circles are the same as in the top panel.}

  \label{images}
\end{figure}

We have observed the PWN \src\  with \xmm\
(\citealt{jla01}) and \suz\  (\citealt{mbi07}) in 2006 and 2007
respectively. Table \ref{obs} summarizes the observations used in this
work. The data have been analyzed with the latest software available,
namely SAS v8.0 for \xmm\ and the pipeline v.2.1.6 with HEASOFT v6.6
for \suz. The \xmm\ data have been screened for proton flares using the
sigma clipping algorithm described in \citet{sk07}, while for \suz\  we
have used the standard screening. The image of \src\  obtained with the
\epic\ PN and MOS CCD cameras of \xmm\  (\citealt{taa01,sbd01}, {
with a resolution of $8^{\prime\prime}$ HPD)} and the XIS CCD camera of
\suz\ (\citealt{ktd07}, { $110^{\prime\prime}$ FWHM}) are shown in
Fig. \ref{images}, in the upper and lower panel respectively.

There is a diffuse extended emission surrounding the bright plerion in
both images. This diffuse emission seems to have spatial structures
as small as $1\arcmin$. The contrast of surface brightness
between the diffuse emission and the core of the PWN is
about 1000.

\begin{table}
  \caption{\suz\  spectral fitting results}
\label{spfits}
\medskip
\centering\begin{minipage}{8.0cm}
\begin{tabular}{lccccc} \hline
  Model      & $N_H$     &  kT or $\gamma$\footnote{the power-law photon index} & $\chi^2/dof$ \\
  & $10^{22}$ cm$^{-2}$ &  keV   &  \\ \hline
  \multicolumn{4}{c}{ { Core of the PWN\footnote{Unabsorbed flux in the 2--10 keV band is $7.5\pm 0.5 \times 10^{-12}$ erg cm$^{-2}$ s$^{-1}$.}}}  \\
  Power-law  & $1.57\pm 0.06$  &  $1.82\pm 0.04$ & 925/1145 \\
  \multicolumn{4}{c}{{ Shell\footnote{Extraction region 91 arcmin$^2$. Unabsorbed flux in the 2--10 keV band is $4.7\pm 0.7 \times 10^{-12}$ erg cm$^{-2}$ s$^{-1}$ without the contribution scattered from the core.}}}  \\
  Power-law  & 1.57  &  $2.9\pm 0.3$ & 173/215    \\
  Thermal    & 1.57  &  $2.0\pm 0.4$  & 196/215     \\

\hline
\end{tabular}
\end{minipage}
\end{table}

\begin{figure}
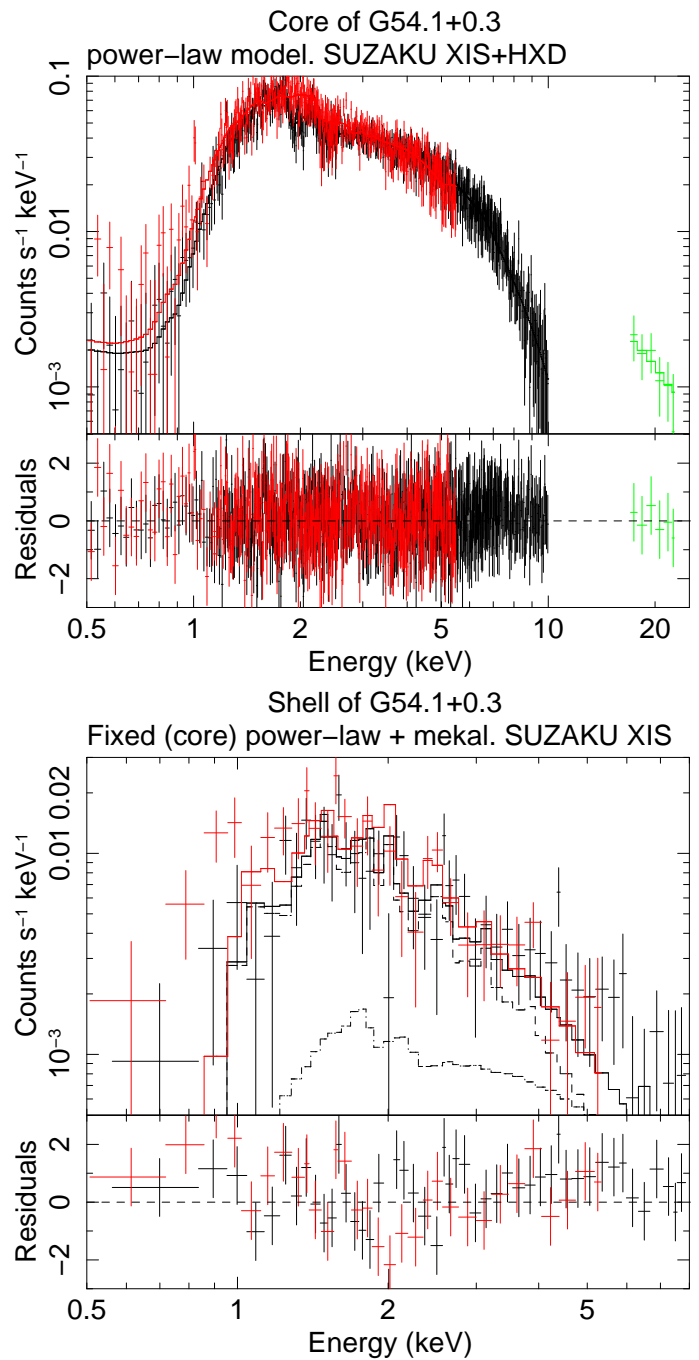

  \centerline{\psfig{file=FIGURE/core_pha.ps,width=9.0cm,angle=-90}}
  \centerline{\psfig{file=FIGURE/suzshell.ps,width=9.0cm,angle=-90}}
  \caption{{\em Top:} \suz\  spectrum of the core region and the best-fit
  model. XIS0+3, XIS2 and HXD spectra are shown. {\em Bottom:} \suz\
  XIS0+3 and XIS2 spectra of the shell region. The total best fit thermal +
  residual non-thermal component from the core is overplotted, as long
  as the individual components (dashed). }

  \label{suzaku}
\end{figure}

We investigated the nature of the X-ray faint diffuse emission by
studying its spectrum both with \xmm\  and \suz. For the spectral
analysis purposes, we defined a core region and a shell region. The
core region is a circle centered on the pulsar with a radius of $160\arcsec$.
This includes all the regions occupied by the radio nebula
and its X-ray emission is dominated by the synchrotron emission coming
form the PWN. The shell region is an annulus with a inner radius equal
to the radius of the core region and an outer radius of $400\arcsec$. We
estimated that the contribution of the core emission in the shell region
is 25\% for \suz\ and 1\% for \xmm, purely due to the instrumental
PSF wings and excluding the additional contribution of dust scattered
X-rays. In the case of \xmm, the shell region was further reduced to
a pie sector between polar angles $90^\circ$ and $135^\circ$ (from N,
anti-clockwise, where most of the knots are located. The background
was taken from the same chip for \suz, while in the case of \xmm\
we have used both blank fields and the same observation to collect
background (in the latter case an annular region with $R_{min}=400\arcsec$
and $R_{max}=450\arcsec$ was used), verifying that the results did not change
with the particular choice of the background. We fitted the core region,
finding that a power-law model describes the data very well. Therefore,
we fitted the shell region using a combination of the power-law model
used in the core (with parameters fixed to their best-fit core values,
including interstellar absorption) and an additional component, chosen
among a thermal and a non-thermal model. The core model was used (and
rescaled) in the shell region to take into account possible contamination
from dust-scattering and from instrumental Point Spread Function. The
\suz\  results are summarized in Table \ref{spfits}, and seem to indicate
that the shell emission which can be modeled either with a thermal
component or with a non-thermal power law in the shell spectrum. \xmm\
gives similar results. The \suz\ spectrum of the core and the shell are
reported in Fig. \ref{suzaku} along with their best-fit models. { The
core is also detected between 15 and 25 keV using the non-imaging \suz\
Hard X-ray Detector (HXD) silicon PIN diodes (spectrum also reported in
Fig. \ref{spfits}, upper panel), with a flux of $\sim 4\times 10^{-12}$
erg cm$^{-2}$ s$^{-1}$ in this band. The normalization constant between
the PIN and the XIS CCD spectra} is $1.3(1.0-1.6)$, a range which includes
the expected value of 1.15 (\citealt{kmt07}).

\section{Removal of dust-scattering halo}

The presence of a foreground medium may affect in more ways the observed
emission from an X-ray source. The best modeled effect is photoelectric
absorption, but also dust scattering of the X-ray photons may be important.
Its produces an apparent halo around the intrinsic source, which is more
prominent at lower energies and may hamper considerably both spectral mapping
analysis of diffuse sources and searches of faint surrounding features.
Scattering halos may be relevant whenever the column density of the intervening
material ($N_H$) is high.
$N_H$ may be derived by fitting the photoelectric absorption of the X-ray
spectrum, while the scattering optical depth ($\tausca$, estimated at a
reference photon energy of 1~keV) by fitting the halo. \citet{ps95} show
that the linear regression
\begin{equation}
  \tausca(1\U{keV})=0.05(N_H/10^{21}\U{cm^{-2}})-0.083
\end{equation}
can be drawn between these two quantities.
For the measured value of $1.57\E{22}\U{cm^{-2}}$ for G54.1+0.3, this
relation predicts $\tausca(1\U{keV})\simeq0.7$.
This value is however only approximate, because the properties of the dust
may be different along different directions, while for the exact value a
direct analysis of the halo is required.

While halos of strong, point-like X-ray sources are selected to investigate
properties of the dust grains, in this case we have to deal with a fainter,
diffuse intrinsic source and therefore the modeling of the halo is necessarily
less accurate. On the other hand, our goal here is simply to find a modeling
that can subtract efficiently the halo component, without pretending to infer
reliable physical properties of the dust distribution.

In the case of a point-like source, we can describe the halo with the
following function:
\begin{equation}
  \Isca(\tht,E)=F(E)\left(1-\exp(-\tausca(E)\right)H(\tht,E).
\end{equation}
where $F(E)$ is the source intrinsic flux and $H(\tht,E)$ is a function that
we derive and describe in detail in the Appendix \ref{app1}, where we also
summarize the basic modeling of a halo and the related assumptions.

In our case, however, the intrinsic source is spatially resolved, and in
principle we should convolve the halo for a point-like source with the
surface brightness distribution of the actual intrinsic source, in a way to
obtain the observed map. This is in general a very complex and numerically
heavy task. Moreover, if just one energy band is considered, the number of
possible solutions would even be infinite. To solve the problem of separating
intrinsic source and halo, one must then carry on a combined fit on radial
profiles at different energies, by taking advantage of the known energy
dependence of the halo properties. In order to model the energy dependence
of the scattering halo, we have produced for each instrument 4 images, in
the following spectral bands: 1.0--1.75~keV, 1.75--2.5~keV, 2.5--3.9~keV,
and 3.9--7.25~keV. The spectral boundaries have been chosen to get similar
numbers of photons in the various bands, with the further constraint of
excluding photons softer than 1~keV, for which the simple Rayleigh-Gans
scalings with the photon energy (namely optical depth $\propto E^{-2}$ and
halo size $\propto E^{-1}$) are no longer valid. The reference energies of
the four bands are 1.4, 2.05, 3.07, and 4.94~keV respectively, and are
obtained averaging the energies of all photons collected in each band.

Since the intrinsic source is centrally peaked and more concentrated than
the halo, and the wings of XMM-Newton PSF are narrower than the observed
radial profile, we have applied a simplified approach, by approximating
the halo with that for a point-like source. For our final fits we have
used only 2 free parameters to model the halo, namely $\tausca(1\U{keV})$
and $\thtscal(1\U{keV})$.  Before then, we had attempted also fits using
a larger number of parameters, but with the moderate statistics of our
data we have found: i. a partial degeneracy between power-law index of
the grain size distribution ($q$) and the spatial scale ($\thtscal$),
and therefore we have chosen a rather usual value for $q$ (3.5, see
e.g.\ \citealt{ps95}); ii. our fits were typically consistent with
a wide spread of $z$ (i.e.\ the position along the line of sight,
normalized to the source distance), and therefore we have decided
to assume a homogeneous distribution of dust along the line of sight
($\zmin=0$ and $\zmax=1$). All quantities cited here are described in
the Appendix \ref{app1}.

For the intrinsic source we have chosen the following modeling:
\begin{equation}
  \Iintr(\tht,E)=\frac{F(E)}{C}\left(\exp\left(-\frac{\tht^2}{\sg_1^2}\right)+A\left(
  1+\frac{\tht^2}{\sg_2^2}\right)^{-\al_2}\right),
\end{equation}
where $C=2\sg_1^2+A\sg_2^2/(\pi(\al_2-1))$ is the normalization
factor.  The fitted shape is indeed a convolution of the actual
source with the instrumental PSF, but at this level we do not
need to separate them\footnote{ { Eq.\ 3 corresponds to the
PSF analytical description discussed in the EPIC Calibration
status document available in the ESA XMM-Newton Calibration Portal
(http://xmm2.esac.esa.int). However,
our best-fit parameters are in general larger since the PWN is
extended. To the purpose of our halo modeling, we just need an empirical
relation to take into account the PSF+source effect.}}. In addition,
we have assumed that this shape is independent of $E$: indeed, it is
known that the X-ray size of PWNe is slightly decreasing
for increasing $E$, but this is only a minor effect, which we cannot
adequately describe with the available data, and on the other hand does
not affect considerably the results of our fits.

From each image we have extracted a logarithmically spaced radial profile.
In order to minimize the statistical noise, the flux values are
averaged over several points.  In the case of MOS1, due to the absence
of the damaged CCD\#6, data are missing for a region that is relevant
to our purposes, { so we use MOS2 only.  By analyzing the emission
at distances larger than $300\arcsec$ from the source center, we have estimated
the MOS2 background levels for the 4 bands, as about $10^{-4}$ times
the surface brightness in the brightest areas.  Therefore, we did not
apply any correction of the background.} With this, we are confident
that our profiles are usable over a dynamical range close to $10^4$.

\begin{table}
  \caption{X-ray dust scattering halo best-fit parameters for \src}
\label{halofit}
\medskip
\centering\begin{minipage}{8.7cm}
\begin{tabular}{lccc}                             \hline
  Name                 & $0^{\prime\prime}-120^{\prime\prime}$  & { $0^{\prime\prime}-200^{\prime\prime}$  }   \\ \hline
  $\sg_1$              & 6.8\arcsec  & { 6.6\arcsec  } \\
  $A$                  & 0.87  & { 0.97  } \\
  $\sg_2$              & 30.7\arcsec & { 29.2\arcsec } \\
  $\al_2$              & 2.27  & { 2.22  } \\
  $\tausca(1\U{keV})$  & 1.15  & { 1.09  } \\
  $\thtscal(1\U{keV})$ & 15.1\arcmin & { 10.2\arcmin } \\ \hline
\end{tabular}
\end{minipage}
\end{table}

For the fits, we have used at first only the inner $120\arcsec$ of the radial
profiles. This to allow an unbiased analysis of structures that may
appear at larger radii.  { However, we also present the results
obtained in the inner $200^{\prime\prime}$ of the radial profiles. The results of the
fits are presented in Table \ref{halofit}. The results are in general
rather similar between different choice of maximum fitting radius. }

\begin{figure}
\centerline{\psfig{file=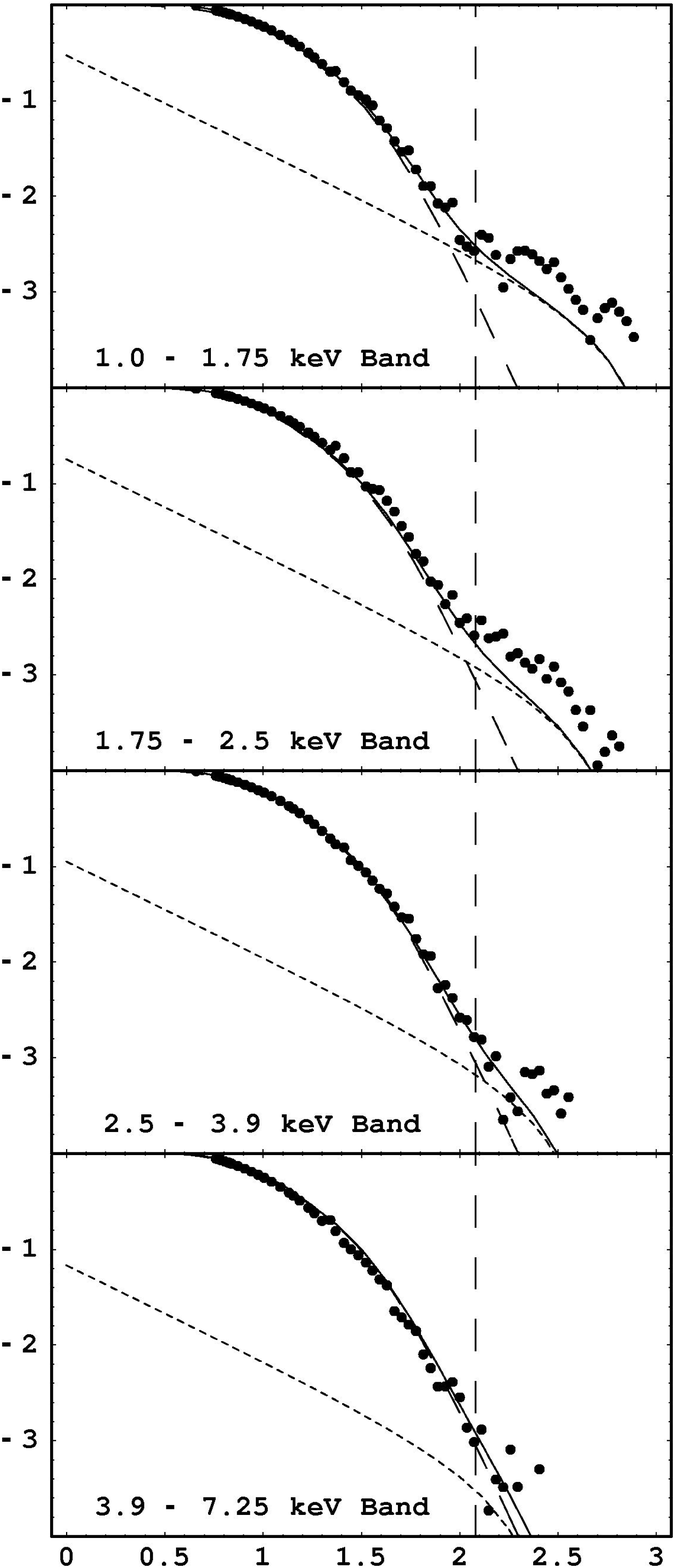,width=8cm}}
  \caption{X-ray normalized profiles in the bands 1.0--1.75, 1.75--2.5,
  2.5--3.9, and 3.9--7.25 keV of \src\  as seen by MOS2 camera of
  \xmm\ \epic. In the x-axis we report the logarithmic distance from
  the center in arcsec. We have overplotted the best fit halo model,
  with individual components (intrinsic source, long-dashed line, and
  halo, short-dashed line). The radial range of the fit extends only to
  $120\arcsec$ (the position of the vertical dashed line). See the text
  for further details on the model. The residuals between $160\arcsec$
  and $400\arcsec$ (2.2 to 2.6 in logarithmic scale) with respect to
  the extrapolation of the halo model are due to the X-ray shell.}

  \label{profile}
\end{figure}

Fig.~ \ref{profile} shows, the combined MOS2 fits on the 4 bands {
(for a maximum fitting radius of $120\arcsec$ marked by the vertical
dashed line) } and the extrapolation of the best-fit profiles to larger
radial distances, compared to the observed profiles.  { The observed
radial profile shows an excess of emission at distances from the center $>
100^{\prime\prime}$ which cannot be explained by the dust-scattering halo.
If we normalize to the fluxes due to the source plus the halo in the
$0^{\prime\prime}-160^{\prime\prime}$ region, the model predicts that the
flux due to the halo in the outer $160^{\prime\prime}-400^{\prime\prime}$
annulus are 0.20, 0.10, 0.04 and 0.01 in the 1.0--1.75, 1.75-2.5, 2.5--3.9
and 3.9-7.25 keV respectively (the flux of the source is 0.01 in the same
region), while the observed values are $0.28\pm0.01$, $0.14\pm0.02$,
$0.11\pm 0.02$, $0.16\pm0.03$, significantly higher than the predicted
halo model. We therefore conclude that the excess is due to intrinsic
emission from the \src\  shell.}

\section{Discussion}

We have reported for the first time the detection of a large faint diffuse
X-ray emission around the PWN \src. We have seen that this excess cannot
be due to the X-ray dust scattering halo, because an accurate modeling
of the halo presented in previous section and in the Appendix A shows
that { the halo model underpredicts the observed emission between
$160\arcsec$ and $400\arcsec$ from the center}, where the shell is
observed. This shell emission has an irregular morphology, but can be
enclosed at most inside a circle of 5.7 arcmin radius centered on the
pulsar position, corresponding to $\sim10.3\,d_{6.2}$~pc.

\begin{figure}
  \centerline{\psfig{file=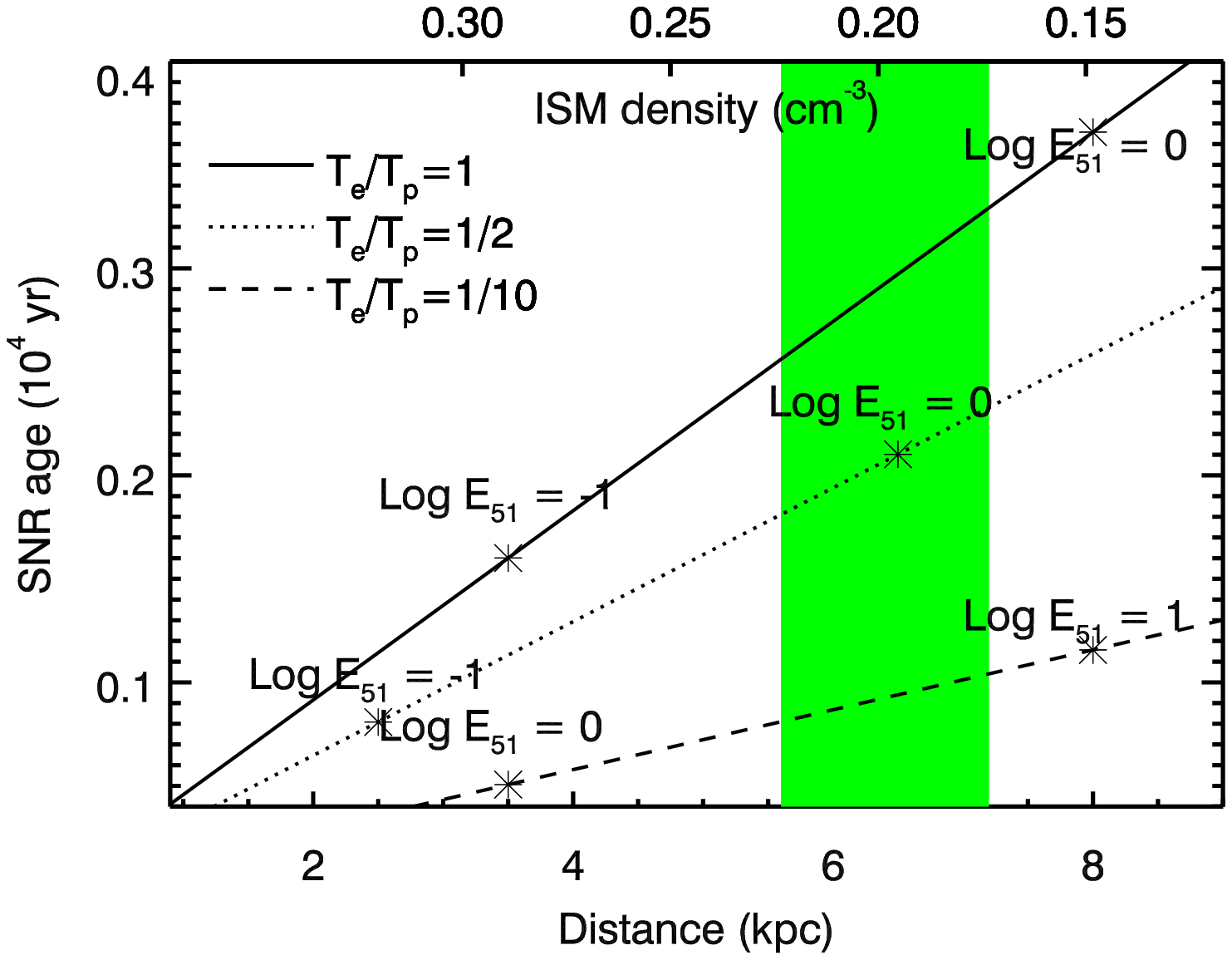,width=9.0cm}}
  \caption{Sedov solutions for the SNR age and distance computed using
  the X-ray derived parameters of \src\  ($kT=2$ keV, SNR radius
  $6\arcmin$, emission measure $9\times 10^{11}$ cm$^{-5}$). We
  report 3 different solutions corresponding to full electron-ion
  equipartition ($T_e/T_p=1$) and the cases of $T_e/T_p=1/2$ and $1/10$.
  The asterisks on the curve mark the corresponding explosion energy of
  the solutions. The intersection of the curves with the area marked
  in green (corresponding to the distance estimate of \citet{ltw08})
  gives the age of the system (for instance, the age is 1800-2400 yr
  for the case $T_e/T_p=1/2$, and the energy is $\sim 10^{51}$ erg).  }

  \label{sedov}
\end{figure}

{ We have seen that the halo spectrum can be interpreted as thermal
or non-thermal emission.  If we interpret it as the long sought thermal
emission of the \src\  shell}, we can derive some interesting quantities
related to the remnant evolution by assuming an expansion governed by the
\citet{Sed59} solution. { The best-fit emission measure is $EM=9\times
10^{-11}$ cm$^{-5}$, and the best-fit temperature is 2 keV. According to
the model of \citet{glr07}, electrons cannot be heated directly at this
temperature by the shock, so the Coulomb heating by the shocked ions must
be at work.  Using the relations (6)--(10) of \citet{bb03c}, and the
best-fit emission measure and temperature, } we have computed the plot
of Fig. \ref{sedov}, which shows all the possible solutions versus the
remnant distance. { We have specifically taken into account the case in
which electrons are not thermalized with ions, by reporting 3 different
cases of electron-ion temperatures in Fig. \ref{sedov}, namely
$T_e/T_p=1$ (electron-ion full equilibration), $T_e/T_p=1/2$ (moderate
electron-ion disequilibrium) and $T_e/T_p=1/10$ (strong disequilibrium, as
in other young SNRs, \citealt{glr07} and references therein)}.  If we use
{ a distance of 6.2 kpc,} we find a range of remnant ages between 2500
and 3300 yr for the equipartition case (solid line in Fig. \ref{sedov}),
which is in agreement with the more uncertain estimate of the remnant age
(1500--6000 yr) given by \citet{clb02}. The inferred ISM pre-shock
density is $\sim 0.2$ cm$^{-3}$ and the swept up mass is between 23
and 32 M$_\odot$, while the X-ray emitting mass is 15--20  M$_\odot$.
The explosion energy range is $E=3-7 \times 10^{50}$ erg. However,
if we drop the assumption of equipartition between electron and ions,
we find that $E=1.0(0.5-1.6)\times 10^{51}$ erg for a distance of 6.2 kpc
when $T_e/T_p \sim 1/2$. In this case, the derived age is between 1800
and 2400 yr (dotted line in Fig. \ref{sedov}). Lower values of $T_e/T_p$
(i.e. $\sim 1/10$) are disfavored by relative high explosion energies
(dashed line in Fig. \ref{sedov}).  We have seen that the XMM spectral
fittings suggest a similar temperature to the \suz\  values, but a
normalization 10 times lower. In this case, the estimate of the ISM
density, explosion energy and swept-up mass must be decreases by a factor
of 3. A cross-check with the non-radiative SNR model of \citet{tm99}
gives a transition from ejecta-dominated to Sedov phase at 2500 yr, so
the remnant is entering in the adiabatic phase.  Given the faintness
of the diffuse emission, only a deeper X-ray observation would allow us
to derive more reliable values of the shell parameters.

The derived values of density and swept-up mass suggest an expansion in
rarefied medium for most of the remnant lifetime. This seems to be in
agreement with the findings of \citet{ltw08} about the environment of
\src. The remnant projected location inside a large IR shell opens up the
possibility that \src\  have been originated by a SN belonging to the same
star cluster whose winds have created the large IR shell. Although the
IR shell distance seems to be a little bit larger than the PWN distance
(7.2 kpc vs 6.2 kpc), according to \citet{ltw08} the uncertainties
on the distance do not rule out the association, and the values of
the density we derived with the X-ray spectral analysis goes in the
same direction. The X-ray shell seems to be larger than the CO cloud
reported by \citet{ltw08}, as shown by the CO contours overplotted in
Fig. \ref{images}.  \citet{kml08} showed that \src\  has interacted
with a star-forming loop located very close to the nebula center (at
$\sim 1\arcmin$ from the pulsar). The loop contains at least 11 young
stellar objects (YSOs) which are very bright in the AKARI 15 $\mu$m
image of the core of the PWN. \citet{kml08} argue that there is no direct
evidence for the interaction of the SNR shock with this dense material,
so they conclude that the IR loops is a partial shell in a low-density
medium and that the SNR shock has propagated well beyond it. This is in
agreement with the position of the X-ray shell we have discovered, since
we can now compute (using the \citealt{tm99} model) that the shock was
at the IR loop position just at 1/10 of the present age. \citet{tsr10}
proposed an explanation in terms of ejecta dust for the IR loop, which
is not in contrast with the presence of the X-ray shell.

\begin{figure}
  \centerline{\psfig{file=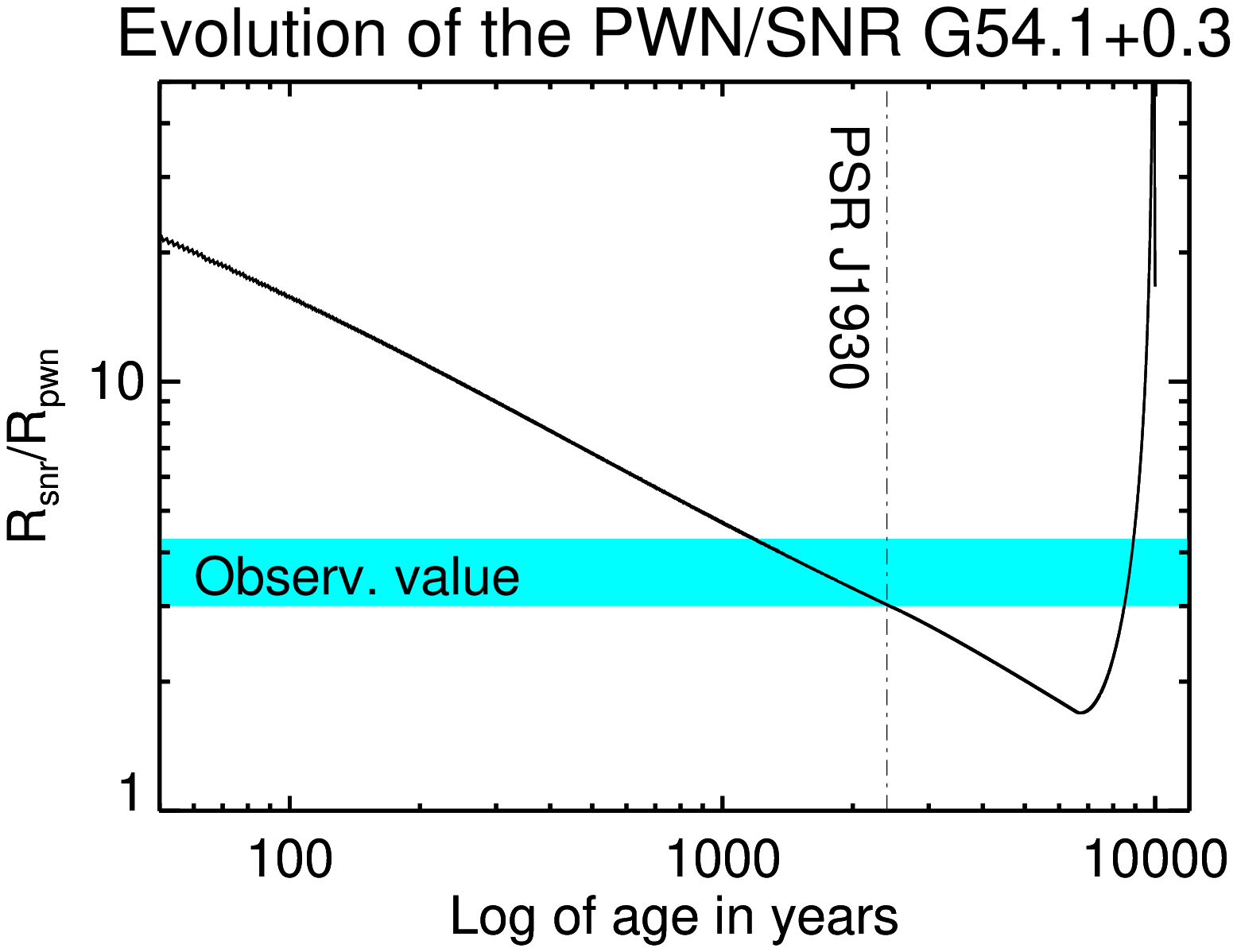,width=9.0cm}}
  \caption{Ratio between the shell radius and PWN radius according to
  the model of \protect\citet{gsz09}, with input parameters tailored to
  the \src\  system (see text for details). The observed value is also
  reported as a cyan stripe, whose spread is due to the uncertainties
  in defining the real PWN radius (we have used a minimum of $80\arcsec$
  and a maximum of $114\arcsec$, on the base of the morphology shown
  in Fig. \protect\ref{images}). The vertical dot-dashed line marks the
  instant of the run when the simulated pulsar has a characteristic age
  and spin-down luminosity identical to PSR J1930+1852 as measured by
  \protect\citet{clb02}. In this moment, the model reproduces both the
  PWN and SNR size and their ratio, strengthening the conclusion that the
  extended diffuse emission is indeed the SNR shell of the PWN \src.  }

  \label{yosi}
\end{figure}

If the faint diffuse emission we have discovered around the PWN \src\
is really the associated SNR shell, then we can compare its properties
with the PWN-SNR evolutionary model of \citet{gsz09}. This model couples
the dynamical and radiative evolution of the PWN with the dynamical
properties of the surrounding non-radiative SNR, and predicts several
distinctive evolutionary stages, namely the initial expansion, the reverse
shock collision, the re-expansion and the second compression.  We run
the model using as input parameters $E_{sn,51}=1$ and $n_{ism}=0.2$
cm$^{-3}$, derived from the best-fit thermal model of the X-ray shell
(Fig. \ref{sedov}, $T_e/T_p=1/2$ case of a moderate deviation
from electron-ion equipartition).  We assumed an ejecta mass of 8
$M_\odot$, and a spin-down timescale of 500 years (close to the value
of the Crab; this corresponds to an initial period $P_0=56$ ms of
the pulsar).  The pulsar wind properties are the same as in Table
2 of \citet{gsz09}.  The resulting dynamical evolution of the
shock of the shell and the PWN nebula is shown in Fig. \ref{yosi},
where we plot the ratio between the shell radius and the PWN radius
($R_{snr}/R_{pwn}$) versus time, and a vertical line marks the
time when the pulsar has the same characteristic age as measured
by \citet{clb02}, and its period and period-derivative match what
was observed ($\tau_c=2900$ yr and a real age of 2400 yr). The predicted
$R_{snr}/R_{pwn}$ is remarkably similar to the observed value, and the
SNR and PWN sizes (8.7 and 2.8 pc) are in relatively good agreement with
observations\footnote{We have verified that, by running the PWN-SNR
model using the results of the equipartition case ($T_e/T_p=1$ in
Fig. \ref{sedov}) we would not reproduce the observed size of the remnant,
unless a far too low ejecta mass is used.}. According to this model, the
PWN has not yet been crushed by the reverse shock (Fig. \ref{yosi} shows
that it will happen at an age of $\sim 7000$ yr, when $R_{snr}/R_{pwn}$
reaches a minimum), in agreement with the lack of sign of crushing,
as noted by \citet{tsr10}.  We conclude that the thermal parameters we
have measured in the faint diffuse emission around the PWN and its size
are in good agreement with what we expect from a putative shell of the
SNR \src, on the basis of a complete modeling of the SNR-PWN system.
We interpret this as a further indication that what we observe is indeed
the long sought shell of the remnant.

\section{Summary and conclusions}

We have analyzed an \xmm\  and a \suz\  observation of the PWN
\src, in the framework of a program aimed to survey the
region around this isolated nebula in search for the X-ray shell of the
associated supernova remnant. We detected very faint X-ray emission
around the PWN, extending form the outskirts of the PWN (at $\sim
1.5^\prime$ from the central pulsar) until a radius $\sim 3.8$ times
the PWN radius (i.e. $\sim 5.7^\prime$, around 10.3 pc at the distance
of the nebula). This extended diffuse emission is more evident toward
south and it has an irregular morphology on a angular scale of $\sim
1^\prime$. We modeled the X-ray dust scattering halo around \src, and
we have found that the detected faint diffuse emission cannot be due to
this effect, but it must be intrinsic to the source.  We modeled the X-ray
spectrum of the diffuse emission with a thermal model, finding a best-fit
temperature of $\sim 2$ keV, { which may imply electron heating
by the shocked ions}. This value, together with the apparent size
and the emission measure of the X-ray emitting plasma, is consistent
with a SNR shell expanding into a $\sim 0.2$ cm$^{-3}$ ISM, whose
explosion energy is $\sim 10^{51}$ erg, { and whose most probable age
is 1800-2400 yr, a bit less than the characteristic age of the pulsar
PSR J1930+1852}, located at the center of the PWN. However, due to
limited counting statistics, the X-ray spectrum of the diffuse emission
can be { alternatively well fitted} with a non-thermal power-law
model, whose photon index ($\gamma = 2.9$) is roughly consistent with
an interpretation in terms of synchrotron emission from accelerated
particles.  The morphology of the large diffuse emission neither seems to be
directly linked to the IR shell observed around the PWN by \citet{kml08},
nor to the molecular cloud detected by \citet{ltw08} and reported
as contours in Fig. \ref{images}, but the fact that the X-ray shell is
incomplete is probably related to the interaction between the PWN and
these inhomogeneities of the ISM.

We have compared the PWN and SNR sizes with the prediction of the
evolutionary model of \citet{gsz09} for composite SNRs, and we find an
excellent agreement. We conclude that the faint diffuse emission around
the PWN \src\  may indeed be the shell of the associated remnant.
However, { deeper X-ray and radio observations are required to
definitely distinguish between thermal and non-thermal interpretation. }
Given the recent detections of X-ray shells around other PWNe, our results
suggest that the lack of shell around remaining isolated PWNe may be
simply the result of high absorption and/or lack of long observations,
and that the X-ray band may be very effective in discovering them.

\begin{acknowledgements}

We thank Prof. D. Leahy and Dr. Wenwu Tien for providing us with the
electronic version of Figure 3 of their work \citet{ltw08}. JDG is
supported by an NSF Astronomy and Astrophysics Postdoctoral Fellowship
under award AST-0702957. This work is partially supported by the ASI-INAF
contract I/088/06/0.

\end{acknowledgements}

\appendix

\section{An approximated treatment of dust-scattering X-ray halos}
\label{app1}

\CHANGED{We assume that scatterings are in} Rayleigh-Gans regime (this is
typically valid above $1\U{keV}$; see, e.g., \citet{sd98} for more
details on the different scattering regimes).  In addition, \CHANGED{we assume that scattering angles are small,}
and \CHANGED{we do} not consider multiple scatterings (e.g.,
\citealt{pk96}).  In this way, one can take advantage of two simple
scaling laws: the scattering optical depth scales with the photon energy
as $E^{-2}$, while the angular scale of the halo scales as $E^{-1}$.

The radial profile of the halo of a point-like source, at a given photon
energy, is derived by calculating the following integral
\begin{equation}
  \Isca(\tht,E)=F(E)N_H\!\!\int \frac{dx}{x^2}f(x)\!\int da\,n(a)
    \dsgdOmd(E,a,\phi),
\end{equation}
where $\tht$ is the (angular) radial distance, $F(E)$ is the source intrinsic
flux, $x$ is the normalized distance to the source ($x=1-z$, following the
notation of \citealt{sd98}), $f(x)$ is the normalized distribution of
dust density along the line of sight, and $n(a)$ is \CHANGED{the} normalized distribution
of grain sizes ($a$; here we assume a position-independent shape of this
distribution).
Finally, the scattering angle $\phi$ is equal to $\tht/x$.

The differential cross section for single scattering can be expressed as
\begin{equation}
  \dsgdOmd(E,a,\phi)=Ca^6G(y)^2,
\end{equation}
where $C$ contains the information on the grain composition, $y$ is defined
as $aE\phi/\hbar c$, and
\begin{equation}
  G(y)=\frac{3}{y^3}(\sin(y)-y\cos(y))=1-\frac{y^2}{10}+\ord(y^4).
\end{equation}
Therefore, assuming a constant $f(x)$ between $\xmin$ and $\xmax$ and zero
elsewhere, and a power-law grain size distribution ($n(a)\propto a^{-q}$
up to a maximum size $\amax$, we have
\begin{equation}
  \frac{\Isca}{F(E)N_H}\propto\!\int_{\xmin}^{\xmax}\!\frac{dx}{x^2}
    \int_0^{\amax}\!\!\!\!\!da\,(7-a)a^{6-q}G\!\left(\frac{aE\tht}{\hbar cx}\right)^2\!\!\!.
\end{equation}
In the limit $\amax\ll\hbar cx/E\tht$, the integral in $a$ is equal to:
\begin{equation}
  \amax^{7-q},
\end{equation}
while, in the opposite limit, it is equal to:
\begin{equation}
   \left(\frac{\xi\,\hbar cx}{E\tht}\right)^{7-q}
  =\left(\amax x\frac{\thtscal}{\tht}\right)^{7-q},
\end{equation}
where we have defined $\thtscal=\xi\,\hbar c/\amax E$ and
\begin{equation}
  \xi^{7-q}=\int_0^\infty dy\,(7-q)y^{6-q}G(y)^2.
\end{equation}
The quantity $\xi$ is a function of $q$ only: it evaluates 2.418 for $q=4$,
2.727 for $q=3.5$, while it slowly diverges for $q$ approaching 3.
Just as an orientative value, for $q=3.5$ and $\amax=0.25\U{\mu m}$ 
(\citealt{sd98}) \CHANGED{we have} $\thtscal\simeq7.4E\rs{keV}^{-1}\U{arcmin}$.

We introduce a ``step-like'' approximation for the function $G(y)^2$: namely,
equal to unity for $y<\xi$ and vanishing elsewhere.
This approximation is equivalent to approximate the integral in $a$ by
matching its two limits.
In this way, it is possible to integrate analytically the integral in Eq.~A.7.
The result is proportional to function $H(\tht,E)$ whose shape defined as 
\begin{equation}
\begin{array}{l l}
  (6-q)(\xmin^{-1}-\xmax^{-1})
   & \mbox{\rm if }\thtn<\xmin		\\
  (7-q)\thtn^{-1}-\xmin^{6-q}\thtn^{-(7-q)}
   & \\
  \qquad-(6-q)\xmax^{-1}
   & \mbox{\rm if }\xmin<\thtn<\xmax	\\
  (\xmax^{6-q}-\xmin^{6-q})\thtn^{-(7-q)}
   & \mbox{\rm if }\xmax<\thtn
\end{array}
\end{equation}
where $\thtn=\tht/\thtscal$.
The total intensity of the profile as defined by Eq.~A.7 is
\begin{equation}
  W=\frac{(6-q)(7-q)}{(5-q)}\left(\xmax-\xmin\right)\pi\thtscal^2.
\end{equation}
It is then convenient to redefine $H(\tht)$ as divided by $W$, so to normalize
its integral.
In this way, we can finally simply write:
\begin{equation}
  \Isca(\tht,E)=F(E)\left(1-\exp(\tausca(E)\right)H(\tht,E).
\end{equation}
With respect to the profile derived here, an exact solution would give a
slightly smoother profile.
However, this effect is minor compared to the uncertainties related to how
sharp is the upper cutoff in the distribution with size and how sharp are
the boundaries of the spatial distribution of grains.

It is worth noticing a few properties of $H(\tht,E)$.
As expected, $\thtscal\propto E^{-1}$; also $\thtscal\propto \amax^{-1}$.
In addition, if $\xmin$ and $\xmax$ are both changed by a factor $\eta$,
the shape of $H$ does not change, provided that $\thtscal$ changes as $\eta$.
This, for instance, implies that the effect of dust extending from the source
to a minimum normalized distance ($\zmin$) to the observer is equivalent
to the case of a uniform spatial distribution of the dust, but with a size
distribution extending to $\amax/\zmin$ instead of to $\amax$: this effect
would lead to overestimate the value of $\amax$.

Since here we do not want to study the actual properties of the foreground
dust but simply to model the shape of the scattering halo, without loss of
generality, in the following we assume $\zmin=0$ (i.e., $\xmax=1$).

\bibliographystyle{aa}
\bibliography{references}

\end{document}